# Heat Transfer Rate Measurements in a Shock-Focused Region in Air

Saranyamol V. S.　　Jithin Sreekumar　　Mohammed Ibrahim S.[1]

Hypersonic Experimental Aerodynamics Laboratory (HEAL)

Department of Aerospace Engineering,

Indian Institute of Technology, Kanpur, U. P., India, 208016.

**ABSTRACT**

An experimental investigation was carried out to study heat transfer rates in a high-temperature, high-pressure region generated using the shock focusing technique. A shock tube test facility with a specially designed spherically converging test section was used in the present study. Two test cases, a shock of initial strength Mach 2 and Mach 4, were investigated. An in-house developed K -type thermocouple was used in the present investigations, and the measured heat transfer rates were of the order of $KW/cm^2$.

## 1  INTRODUCTION

The phenomenon of focusing a high-energy shock wave to a small region in space resulting in high energy concentration, is called shock wave focusing. There are several ways to achieve shock wave focusing: like using a shock tube[1], [2], exploding wire [3], micro explosive [4], [5], electrical discharge [6], lasers [7], etc. Among these, the most common and easy way is to produce a shock wave with the help of a shock tube and focus it with various techniques. There are several applications to the phenomenon of shock wave focusing. Inertial confinement fusion [8][9], shock wave lithotripsy, material science [3], ignition techniques [4], etc., are some of them.

The research based on shock focusing phenomenon has started since 1942 when Guderley [12] did theoretical studies on strong converging cylindrical and spherical shock waves. He proposed a self-similar solution in an ideal gas flow, which express the radius of a converging shock wave as a function of time. Pioneer experimental study on converging shock waves was done by Perry and Kantrowitz [11], achieving cylindrical shock focusing with a teardrop insert placed inside a shock tube. Detailed characterization and study on cylindrical shock focusing in a shock tube was carried out by Zhai et al. [14], [15]. Several other shock focusing techniques, like parabolic reflector in shock tube [1], annular shock tube [16], were also achieved by researchers.

---

[1] Corresponding author: *ibrahim@iitk.ac.in*

Spherical shock waves are capable of focusing more energy and are a more challenging method compared to cylindrical shocks [17], [18]. With a shock of the same incident shock strength, spherical convergence can accelerate it to higher Mach number values when compared to that of cylindrical convergence [18]. Pioneering experiments on spherical shock wave focusing were carried out by Roberts and Glass [19] using a hemispherical implosion chamber. A blast wave is produced by reflecting a detonation wave generated by the explosion of the $O_2$-$H_2$-He mixture. Achieving spherical shock wave focusing in a shock tube is a bit challenging due to the concern about the stability of converging shock. Spherical shock wave focusing in a shock tube was achieved by Setchell et al. [20] with the help of a 10º half-angle cone attached to the driven end of the shock tube. This procedure of focusing the shock was associated with several Mach reflections, resulting in losses to the shock front. The losses were overcome by Saillard et al. [21] by passing the shock through a smooth curve. The challenge of focusing a spherical shock with minimum losses was also successfully overcome by Liverts and Apazidis [17], Kjellander et al. [22][23], and Liverts et al.[24]. A perfectly contoured converging section helped to smoothly vary the shock profile from planar to spherical with minimum diffusion losses to the shock front [25], [26]. The shock front accelerates while passing through a converging section. The temperature at the focusing region reaches high enough that the gas in this region starts radiating [27]. Several researchers have captured this radiation, and the temperature of this radiating flow field is obtained by comparing the radiating spectrum to that of a black body curve [28].

The temperature at the focusing region jumps drastically, that the heat transfer rate shoots to high values at these regions. There are several ways to measure the heat transfer rate at a point. Under the assumption of a one-dimensional semi-infinite heat conduction equation, the transient temperature history is typically processed and converted into heat flux using well-established methodologies [29]. Various sensors are used to obtain this, such as platinum thin film gauges [30]–[33], coaxial thermocouples [34]–[36], thin skin thermocouples [37], and null point calorimeter [32], [38]. However, the heat transfer measurement for an application like shock wave focusing has not been attempted by any researchers yet.

In the current study, an attempt is made to directly measure the heat transfer rate at the focusing region using a spherically converging test section attached to a shock tube test facility.

Experiments were performed at two different initial shock strengths. The heat transfer measurement is carried out by in-house made coaxial thermocouples. Fast-response K-type thermocouples are designed, fabricated, and calibrated. An adaptor is designed to house the thermocouple to the focusing region of the converging section. This gives the temperature rise through which the heat transfer can be estimated.

## 2 EXPERIMENTAL METHODOLOGY

### 2.1 Shock Focusing Facility

Experiments are performed using the shock tube facility, 'S1 (Vaigai)' at the Hypersonic Experimental Aerodynamics Laboratory (HEAL), Indian Institute of Technology, Kanpur, India. The facility has a 1 m long driver section, a 7 m long driven section, and an 87 mm internal diameter. An aluminium diaphragm separates the driver and driven sections. A converging section of length 0.3 m is attached at the driven section end, which helps to focus the shock wave. The converging section transforms the planar shock generated in the shock tube into a spherical shock. The converging section is designed in such a way that the wall is smoothly curved so that the shock front will have the minimum possible diffraction losses. The inner wall of the converging section remains perpendicular to the base of the shock, which in turn minimizes the chance of Mach reflections and reattachments happening to the shock front. The internal diameter of the tube reduces smoothly from 87 mm to 18 mm at the focusing end wall by the converging section. An additional converging cone is inserted here, where the diameter further reduces to 4 mm. The thermocouple is fixed in this location with the help of Teflon windings, normal to the flow. The design of the converging section was made according to the geometric relations mentioned in equation 1, which was adapted from Kjellander's work [39].

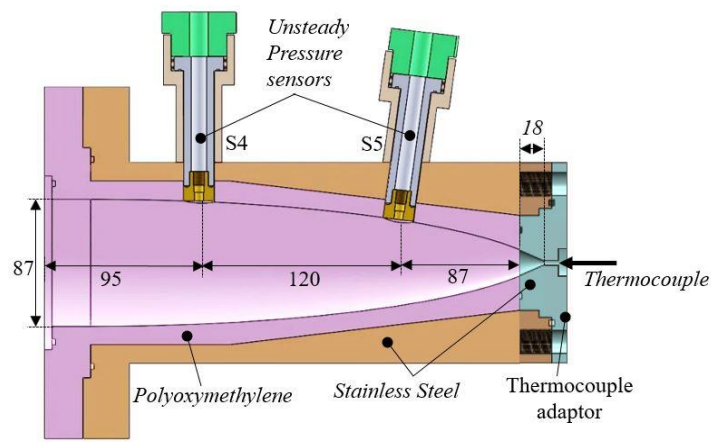

*Figure 1: Converging section with thermocouple adaptor*

$$x = A \sin \theta$$
$$y = B - R(1 - \cos \theta)$$
(1)

where $\theta$ is ranging from 0 to $0.35\pi$; A= 339.7 mm; B= 43.5 mm, and R= 63 mm.

The converging section is designed separately with Polyoxymethylene (Delrin) material and is used as an insert that fits inside the SS outer layer, as shown in Figure 1.

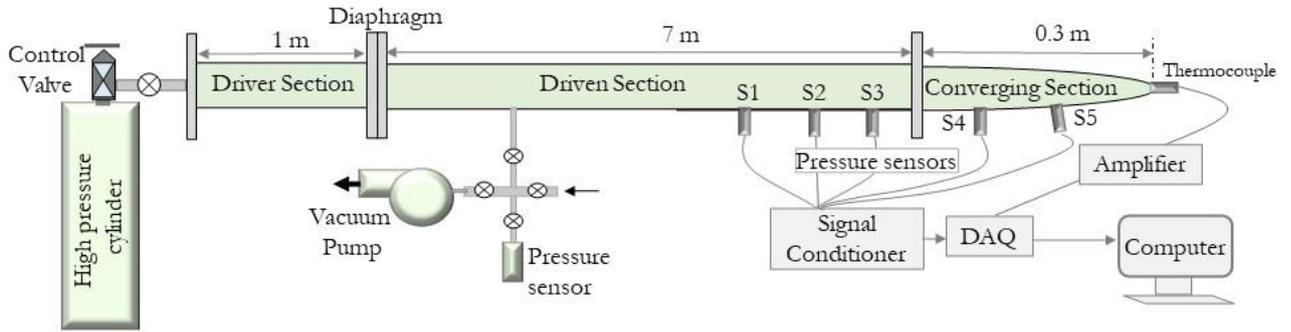

*Figure 2: Schematic diagram of shock focusing setup.*

## 2.2 Instrumentation

Unsteady pressure measurements are carried out using the ICP® pressure sensor of PCB piezotronics, model No-113B22, which are flush-mounted along the surface of the facility, as can be seen in Figure 2. The sensors S1, S2, and S3 are flush mounted on the shock tube, which is used to obtain the initial shock strength. The sensors S4 and S5 are flush mounted on the converging section. The pressure data was acquired using NI-USB 6356, a multifunction I/O DAQ device, at a rate of 1.0 Mega samples per channel over a duration of 0.25 seconds. The DAQ system is triggered and starts acquiring data when the shock reaches sensor S1. The sensors were connected to the DAQ through a PCB signal conditioner (Model No. 482C05).

The K-type thermocouple is made by the Chromel-Alumel combination. The base material, Chromel, is 3.5 mm in diameter, and at the center, Alumel is inserted, as shown in Figure 3a. The in-house-made thermocouple image is shown in Figure 3b. The thermocouples used are 5mm long. K-type thermocouples are used for all the current experiments since they are robust, respond fast, and can be contoured to any type of model surface [40]. K-type thermocouple also has the advantage of higher sensitivity, ~40μV/°C, which can also withhold more test runs. The measurement from the thermocouple is obtained through the data acquisition system. The thermocouple signal is amplified by an amplifier with 1000 gain before connecting to the DAQ system, as seen in Figure 2.

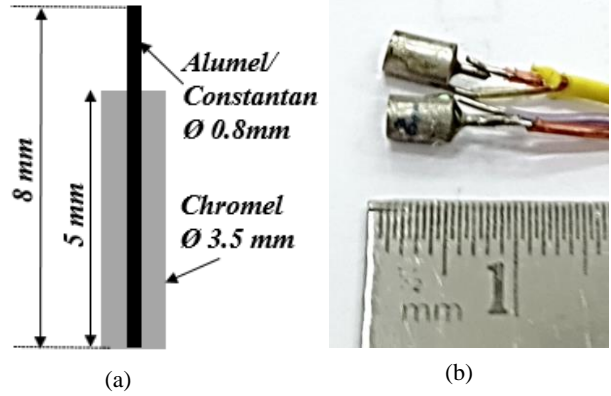

Figure 3: (a) Cross-section of Thermocouple (b) In-house made Thermocouple.

## 2.3 Thermocouple Working Principle

The sensing element on a K-type thermocouple is a thin junction of Alumel inserted in the thermocouple body of Chromel material. The heat flux is obtained from a temperature distribution data of a thermocouple from the one-dimensional heat conduction theory. The thermocouple works based on three assumptions[41], [42]. First, the sensing surface has a negligible effect on the thermal behaviors of the surface. In the case of K-type thermocouple, the Alumel material has no effect on the temperature measured since its diameter (0.8 mm) is less compared to the diameter of the substrate material (3.0 mm), Chromel. The second assumption is that the sensing substrate acts as a semi-infinite solid., and that there is zero temperature difference at infinity. The third assumption is that there is no penetration of heat along the lateral direction. The conduction occurs only in the normal direction. The governing equations for solving a one-Dimensional, semi-infinite, solid body can be expressed by a partial differential equation given in equation 2.

$$\frac{\partial}{\partial x}\left(k\frac{dT}{dx}\right) = \rho C_p \frac{\partial T}{\partial t} \qquad 2$$

Here, k is the thermal conductivity; Cp is the specific heat, and $\rho$ is the density of the substrate, are the thermal properties. Assuming the thermal properties to be constant, the temperature distribution can be given by:

$$\frac{\partial^2 T}{\partial x^2} = \frac{1}{\alpha_0}\frac{\partial T}{\partial t} \qquad 3$$

Were, the thermal diffusivity, $\alpha_0 = \frac{k}{\rho C_p} = constant$.

The boundary conditions to solve the above equation are given by: $T_{(x,t=t_0)} = T_0$, $T_{(x=0,t)} = T_s(t)$, $\dot{q}_{(x=\infty,t)} = 0$.

For high-speed flows, all the thermal properties cannot be assumed to be constant. Hence a better analytical solution is obtained by assuming the thermal conductivity of the substrate as a function of temperature. The thermal diffusivity is assumed to be a constant here. This assumption is reasonable for metallic substrates which are used in thermocouples [41], [42].

The solution is obtained using the Laplace transformation. The solution contains an integral part, which is assumed as a summation to avoid singularity. This assumption is made by taking temperature as a piecewise linear function, as per Cook and Felderman [29]. The total heat added to the substrate can be calculated using the method mentioned by Kendall et al. [43].

$$Q_n = \frac{\beta_0}{\sqrt{\pi}} \sum_{i=1}^{n} \frac{\Phi_i + \Phi_{i-1}}{\sqrt{t_n - t_i} + \sqrt{t_n - t_{i-1}}} \Delta t \qquad 4$$

The heat transfer rate can be computed with the finite-difference approximation used by Hedlund et al. [44].

$$\dot{q}(t_n) = \dot{q}_n = \frac{dQ_n}{dt} = \frac{-2Q_{n-8} - Q_{n-4} + Q_{n+4} + 2Q_{n+8}}{40\,\Delta t} \qquad 5$$

Here,

$$\Phi = \int_{T_0}^{T} \frac{k}{k_0} dT \qquad 6$$

### 2.3.1  *Material properties*

The metal substrate of the thermocouples has high thermal conductivity and thermal diffusivity. This way, when exposed to heat, the temperature diffuses faster, which makes it more suitable for high-speed flow cases. When the material is exposed to high temperatures for a longer time, the assumption of a semi-infinite body is no longer applicable. The material properties of the Chromel [41], [42], which is the substrate of the K-type thermocouple, are mentioned below.

$$\rho = 8714 \ (Kg/m^3)$$
$$c_p = 386.25 + 0.23981\,T \ (J/Kg\,K)$$
$$k = 11.845 + 1.9132 \times 10^{-2} T \ (W/mK) \qquad 7$$
$$\beta = 6398.4 + 6.6331\,T \ (Ws^{0.5}/m^2 K)$$
$$\alpha = 3.5995 \times 10^{-6} + 2.9656 \times 10^{-9} T - 9.1293 \times 10^{-13} T^2 \ (m^2/s)$$

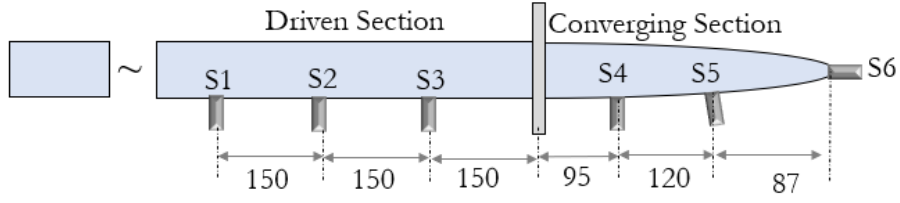

Figure 4: Unsteady pressure sensor locations. All dimensions are in mm.

## 3 RESULTS AND DISCUSSION

Two sets of experiments were carried out in order to obtain the heat transfer measurement showing the effect of shock strength. The test gas used in the driven section for all the tests was Air, and the gas filled in the driver section was helium. The different shock strengths achieved were $M_s$= 2.2 and 4.1. The details of the test conditions are mentioned in Table 1.

In order to get an estimate of the pressure obtained at the focusing region, the thermocouple in Figure 2 was replaced with an unsteady pressure, as seen in Figure 4. The distance between each sensor is also shown here. The pressure distribution at locations S4, S5, and S6 for cases Ms 2.0 and Ms 4.0 are shown in Figure 5.

Table 1: *Test conditions*

| $M_S$ | P1(MPa) | Nomenclature |
|---|---|---|
| 2.2 | 0.025 | $M_S$ 2.0 |
| 4.1 | 0.01 | $M_S$ 4.0 |

The time at which the shock reaches the focusing end wall is taken as zero for comparison purposes. The first rise in the pressure signal on S4 and S5 corresponds to the arrival of the incident shock, and the second rise corresponds to that of the reflected shock. Behind the incident shock, it is seen that the pressure gradually increases until the reflected shock reaches. This is due to an acceleration of the shock front, which results in the strengthening of the shock while moving through the converging section. An abrupt rise in the pressure signal is seen at sensor location S6 due to the shock impact. The incident shock hits the S6 location and reflects as a reflected shock. The fill pressure for case Ms 2.0 is 0.025 MPa in the driven section, which rises to a value of 3.76 MPa due to the shock impact at S6. Similarly, a pressure rise of 6.11 is seen Ms 4.0 case at S6, where the initial fill pressure is 0.01 MPa. The conditions at the focusing region are found to be rather high. Measuring the temperature at this location is challenging since the condition remains only for a few microseconds. Hence the heat transfer occurring in this focusing region is measured with the help of a thermocouple, which obtains the temperature difference.

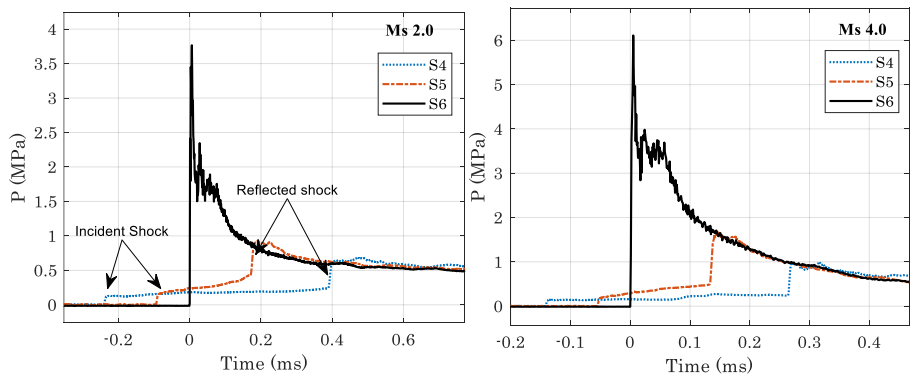

*Figure 5: Pressure measured on the sensors flush mounted on the converging section wall (S4 and S5) and on sensor facing the flow normally (S6) for both Ms 2.0 and Ms 4.0 cases.*

The thermocouple measurement obtained is subjected to several uncertainties. Hence test runs were repeated, and the repeated results for case Ms 2.0 is shown in Figure 6. Test case Ms 2.0 shows good repeatable signals, where all the test runs gave a perfect match with each other. The average standard deviation among the temperatures obtained is 8.5 K.

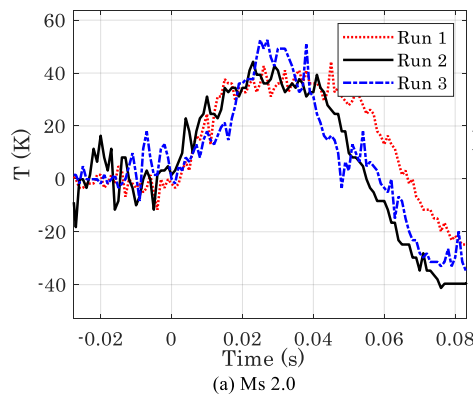

(a) Ms 2.0

*Figure 6: Repeatability of the thermocouple signal*

The temperature distribution measured at the focusing end wall and the heat transfer obtained for case Ms 2.0 is shown in Figure 7. The voltage signal obtained from the thermocouple is converted into the temperature through the sensitivity of the thermocouple, 38μV/°C. For the case Ms 2.0, the temperature rises to 40 K. When the incident shock hits the end wall, the temperature and pressure rise abruptly. As the shock reflects from the end wall, the temperature also drops, which corresponds to the fall, as seen in the figure. The heat transfer rate at the focusing region is obtained from the temperature measurements using the relations mentioned in equation 5. A MATLAB code was developed to generate heat transfer signals. The first rise and fall cycle of the temperature signal was used to obtain the heat transfer. The heat transfer rate measured for cases Ms 2.0 is shown in Figure 7b. The temperature rise caused a maximum heat transfer rate of 7173 W/cm$^2$. The rise time marked in the figure is the time taken for the heat transfer to rise from zero to the peak value. After a rise time of 16 μs, the heat transfer rate remains constant value for the duration of 11µs, which is the same time that it took for the temperature to rise to the maximum value. This duration is marked in both the temperature distribution image and heat transfer distribution image for comparison purposes.

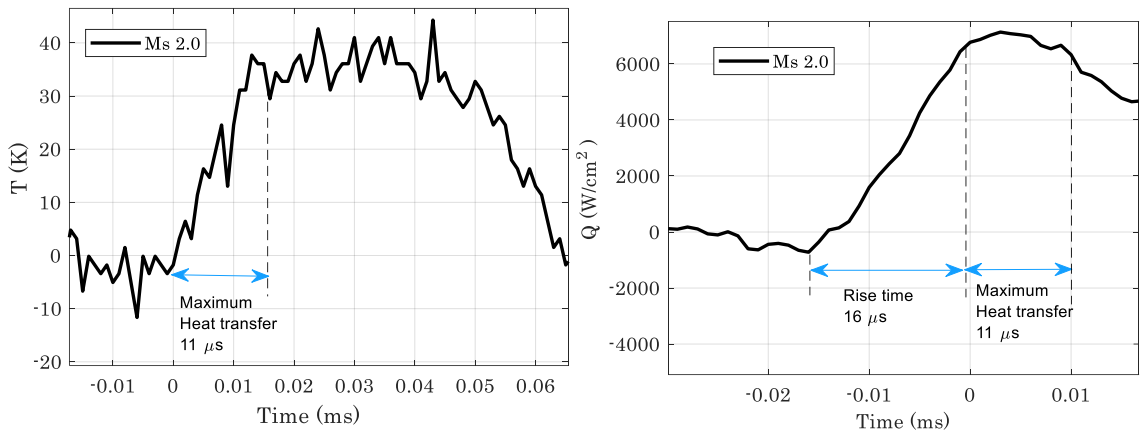

*Figure 7: Temperature measured at the focusing end wall and the heat transfer obtained for Ms 2.0 case.*

The temperature distribution and the heat transfer rate measured for cases Ms 4.0 is shown in Figure 8. For Mach 4.0 case, the temperature rises from zero to 300 K and stays constant for a duration of 5 µs. The magnitude of the temperature increased by increasing the Mach number of incident shock, which resulted in a reduction of fluctuation in the measured signal. The time taken for the temperature distribution to reach a maximum value is 25 µs. The heat transfer value remains constant value of $5 \times 10^4 \ w/cm^2$ for the same duration of time. The rise time has reduced to 12 µs for Ms 4.0 case. Based on repeated tests, the uncertainties (in terms of percentage) in the measured heat transfer are ± 12%.

The thermocouple was connected at the focusing end of the converging section. The conditions of the hot gas at the focusing region have high temperatures and pressure that can cause the gas to dissociate and undergo chemical reactions. Even though K-type thermocouples are well suited for applications where oxidation happens [42], there is a possibility of damage to the thermocouple. The possibility of variation in the sensitivity of the thermocouple due to the impact of the high-strength shock is there. Hence, the thermocouples were calibrated before and after test runs to ensure their working condition. The sensitivity was found to remain constant for thermocouples after repeated tests. The calibration report of the thermocouple before and after the test run is shown in Figure 9. The sensitivity for the thermocouple fabricated was obtained to be 38μV/℃. This demonstrates the robustness of our in-house developed thermocouple sensor, which will be used for future testing in high-speed high-temperature environments.

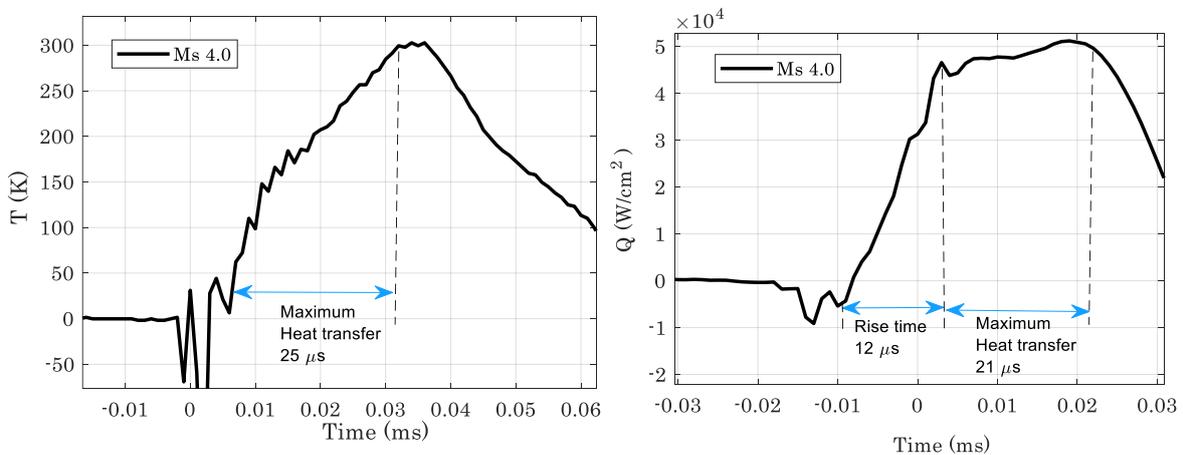

*Figure 8: Temperature measured at the focusing end wall and the heat transfer obtained for Ms 4.0 case.*

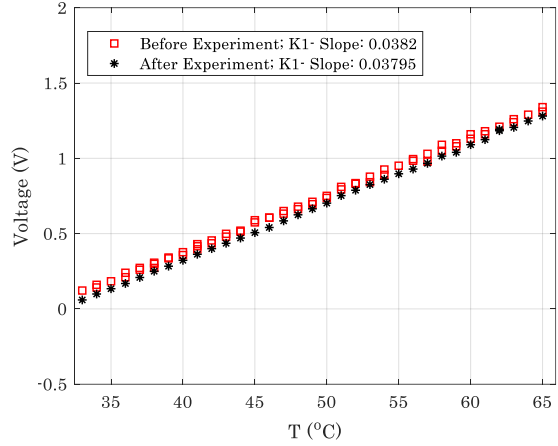

*Figure 9: Sensitivity of thermocouple*

# 4    CONCLUSION

An experimental attempt was made to estimate heat flux in a shock-focused region using a shock tube test facility. A spherically converging test section was designed and used for this study. The study was performed for two test cases of initial shock strength of Mach 2 and 4, and the measured heat flux was around ~7 kW/cm$^2$ and ~50 kW/cm$^2$. Doubling the shock strength increased the heat flux at the focusing region by an order higher. The in-house K type thermocouples used were robust enough to measure the heat flux in such harsh, high pressure, and high temperature environments.

**ACKNOWLEDGMENTS**

The authors would like to thank the Science and Engineering Research Board (SERB), India, for supporting this research work under the Early Career Research Award, ECRA/2018/000678.

# AUTHOR DECLARATIONS

## CONFLICT OF INTEREST

The authors have no conflicts to disclose.

## DATA AVAILABILITY

The data that support the findings of this study are available within the article.